\documentclass{article}
\usepackage{amssymb}

\input{tcilatex}

\begin{document}

\title{Pseudohermitian Hamiltonians, time-reversal invariance and Kramers degeneracy}
\author{G. Scolarici\thanks{%
e-mail: scolarici@le.infn.it} and L. Solombrino\thanks{%
e-mail: solombrino@le.infn.it} \\
%EndAName
Dipartimento di Fisica dell'Universit\`{a} di Lecce \\
and INFN, Sezione di Lecce, I-73100 Lecce, Italy}
\maketitle

\begin{abstract}
A necessary and sufficient condition in order that a (diagonalizable)
pseudohermitian operator admits an antilinear symmetry $\frak{T}$ such that $%
\frak{T}^{2}=-\mathbf{1}$ is proven. This result can be used as a quick test
on the $T$-invariance properties of pseudohermitian Hamiltonians, and such
test is indeed applied, as an example, to the Mashhoon-Papini Hamiltonian.

\end{abstract}

\section{Introduction}

Non Hermitian Hamiltonians are usually taken into account in order to
describe dissipative systems or decay processes. In particular, in the last
few years, a great attention has been devoted to the study of $PT$-symmetric
quantum systems \cite{m6}, whose Hamiltonians (though non Hermitian) possess
real spectra, and in this context the interest rose on the class of
pseudohermitian operators \cite{lee}, i.e., those operators which satisfy 
\begin{equation}
\eta H\eta ^{-1}=H^{\dagger }
\end{equation}
with $\eta =\eta ^{\dagger }$ (of course, Hermiticity constitutes a
particular case of pseudohermiticity, corresponding to $\eta =\mathbf{1}).$

When one considers diagonalizable operators with a discrete spectrum, one
can prove that $H$ is pseudohermitian if and only if its eigenvalues are
either real or come in complex-conjugate pairs (with the same multiplicity) 
\cite{mI}; furthermore, this result has been generalized to all the
(possibly non diagonalizable) matrix Hamiltonians \cite{mV}, and to the
class of all the $PT$-symmetric standard Hamiltonians having $\mathbf{R}$ as
their configuration space \cite{mIV} (which suggests that it may be valid
under more general conditions).

Another physical reason for studying pseudohermitian operators is the remark
that any $T$-invariant (diagonalizable) Hamiltonian must belong to their
class \cite{so}. The converse does not hold in general. Indeed, whereas one
can prove that to any pseudohermitian operator is associated an antilinear
symmetry \cite{so}\cite{mIII}, (in particular, at least, an involutory one),
in general one cannot interpret it as the time-reversal operator $T$;
furthermore, in case of fermionic systems, it is well known that 
\begin{equation}
T^{2}=-\mathbf{1,}
\end{equation}
and the above-mentioned theorems do not ensure the existence of such a
symmetry.

In order to deepen this point, we will prove in Sect. 2 that a Kramers-like
degeneracy is a necessary and sufficient condition so as a diagonalizable
pseudohermitian operator admits an antilinear symmetry which satisfies
condition (2).

Next, as an example, we will apply in Sect. 3 the above result to the study
of a non Hermitian Hamiltonian which has been recently proposed to interpret
(by a $T$-violating spin-rotation coupling) a discrepancy between
experimental and theoretical values of the muon's $g-2$ factor \cite{pa},
and we will able to state precisely the parameters values associated with
the $T$-violation.

\section{A theorem on pseudohermitian operators}

As in \cite{mI}\cite{so}\cite{mIII}, we consider here only diagonalizable
operators $H$ with a discrete spectrum. Then, a complete biorthonormal
eigenbasis $\left\{ \left| \psi _{n,a}\right\rangle ,\left| \phi
_{n,a}\right\rangle \right\} $ exists \cite{fa}, i.e., a basis such that 
\begin{equation}
H\left| \psi _{n,a}\right\rangle =E_{n}\left| \psi _{n,a}\right\rangle
,\qquad H^{\dagger }\left| \phi _{n,a}\right\rangle =E_{n}^{\ast }\left|
\phi _{n,a}\right\rangle ,
\end{equation}
\begin{equation}
\left\langle \phi _{m,b}\right. \left| \psi _{n,a}\right\rangle =\delta
_{mn}\delta _{ab},
\end{equation}
\begin{equation}
\sum_{n}\sum_{a=1}^{d_{n}}\left| \psi _{n,a}\right\rangle \left\langle \phi
_{n,a}\right| =\sum_{n}\sum_{a=1}^{d_{n}}\left| \phi _{n,a}\right\rangle
\left\langle \psi _{n,a}\right| =\mathbf{1,}
\end{equation}
where $a,b$ are degeneracy labels and $d_{n}$ denotes the degeneracy of $%
E_{n}$ ; hence, the operator $H$ can be written in the form 
\begin{equation}
H=\sum_{n}\sum_{a=1}^{d_{n}}\left| \psi _{n,a}\right\rangle
E_{n}\left\langle \phi _{n,a}\right| .  \label{H}
\end{equation}

We can now state the following

\bigskip

\textbf{Theorem.}\textit{\ Let }$H$\textit{\ be diagonalizable operator with
a discrete spectrum. Then, the following conditions are equivalent:}

\textit{i) an antilinear operator }$\mathfrak{T}$\textit{\ exists such that }%
$\left[ H,\mathfrak{T}\right] =0$\textit{\ , with }$\mathfrak{T}^{2}=-1;$

\textit{\bigskip ii) }$H$\textit{\ is pseudohermitian and the degeneracy of
its real eigenvalues is even}.

\bigskip

\textbf{Proof}. Let us assume that condition $i)$ holds; then, $H$ is
pseudohermitian (see \cite{so}, Prop. 3 and Prop.1), hence its eigenvalues
are either real or come in complex-conjugate pairs (with the same
multiplicity). We will use in the following the subscript `$_{0}$' to denote
real eigenvalues and the corresponding eigenvectors, and the subscript `$%
_{\pm }$' to denote the complex eigenvalues with positive or negative
imaginary part, respectively, and the corresponding eigenvectors.

Let now $\left| \psi _{n_{0},a}\right\rangle $ be an eigenvector of $H$;
then, $\mathfrak{T}\left| \psi _{n_{0},a}\right\rangle $ too is an
eigenvector of $H$, corresponding to the same eigenvalue $E_{n_{0}}$, and
linearly independent from $\left| \psi _{n_{0},a}\right\rangle $ . (Indeed,
would be $\mathfrak{T}$ $\left| \psi _{n_{0},a}\right\rangle =\alpha \left|
\psi _{n_{0},a}\right\rangle $ for some $\alpha \in \mathbf{C},$ applying
again $\mathfrak{T}$ to the previous relation we would obtain $\left| \psi
_{n_{0},a}\right\rangle =-\left| \alpha \right| ^{2}\left| \psi
_{n_{0},a}\right\rangle $, which is absurd.)

If $\left| \psi _{n_{0},b}\right\rangle $ is another eigenvector of $H$,
linearly independent from $\left| \psi _{n_{0},a}\right\rangle $ and $%
\mathfrak{T}\left| \psi _{n_{0},a}\right\rangle $, also $\mathfrak{T}\left|
\psi _{n_{0},b}\right\rangle $ is linearly independent from all three,
otherwise, applying once again $\mathfrak{T}$ to the relation 
\[
\alpha \left| \psi _{n_{0},a}\right\rangle +\beta \mathfrak{T}\left| \psi
_{n_{0}.a}\right\rangle +\gamma \left| \psi _{n_{0},b}\right\rangle +\delta %
\mathfrak{T}\left| \psi _{n_{0},b}\right\rangle =0 
\]
we could eliminate, for instance, $\mathfrak{T}\left| \psi
_{n_{0},b}\right\rangle $ obtaining so a linear dependence between $\left|
\psi _{n_{0},a}\right\rangle ,\mathfrak{T}\left| \psi
_{n_{0},a}\right\rangle $ and $\left| \psi _{n_{0},b}\right\rangle $,
contrary to the previous hypothesis.

We can conclude, iterating this procedure, that $d_{n_{0\text{ }}}$must be
necessarily even.

\bigskip

Conversely, let condition $ii)$ hold, and let $\mathfrak{T}$ denote the
following antilinear operator: 
\begin{eqnarray}
\mathfrak{T} &=&\sum_{n_{0}}\sum_{a=1}^{d_{n_{0}}/2}\left( \left| \psi
_{n_{0},a}\right\rangle K\left\langle \phi _{n_{0},a+d_{n_{o}/2}}\right|
-\left| \psi _{n_{0},a+d_{n_{o}/2}}\right\rangle K\left\langle \phi
_{n_{0},a}\right| \right) \\
&&+\sum_{n_{+},n_{-},a}\left( \left| \psi _{n_{-},a}\right\rangle
K\left\langle \phi _{n_{+},a}\right| -\left| \psi _{n_{+},a}\right\rangle
K\left\langle \phi _{n_{-},a}\right| \right) ,  \nonumber
\end{eqnarray}
where the operator $K$ acts transforming each complex number on the right
into its complex-conjugate. Then, one immediately obtains, by inspection,
that $\left[ H,\mathfrak{T}\right] =0$ and $\mathfrak{T}^{2}=-\mathbf{1}.$ $%
\blacksquare $

\bigskip

The implication $i)\Longrightarrow ii)$ we proven above generalizes from
various point of view the celebrated Kramers theorem on the degeneracy of
any fermionic (Hermitian) Hamiltonian. Indeed, it applies to a larger class
than that of the Hermitian operators (concerning their real eigenvalues
only); moreover, it does not require a physical interpretation of the
antilinear operator $\mathfrak{T}$ as a time-reversal operator. However, by
an abuse of language, we will continue to denote as ''\textit{Kramers
degeneracy}'' this feature of pseudohermitian operators admitting a symmetry
like $\mathfrak{T.}$

We stress once more that the Kramers degeneracy is a necessary but not a
sufficient condition for the $T$-invariance.

\section{\protect\bigskip Time-reversal violation in the spin-rotation
coupling}

On the basis of the previous discussions, we can quickly test the $T$%
-invariance properties of pseudohermitian Hamiltonians. To illustrate this
point with an example, we chose a pseudohermitian Hamiltonian which has been
recently introduced to interpret a discrepancy between experimental and
standard model values of the muon's anomalous $g$ factor.

In this model, a spin-rotation coupling, which involves small violations of
the conservation of $P$ and $T$, is considered. In particular, the
spin-rotation effect described by Mashhoon \cite{ma} attributes an energy $-%
\frac{\hslash }{2}\overrightarrow{\omega }.\overrightarrow{\sigma }$ to a
spin-$\frac{1}{2}$ particle in a frame rotating with angular velocity $%
\omega $ relative to an inertial frame. In the modified Mashhoon model \cite
{pa} one assumes a different coupling of rotation to the right and left
helicity states\ of the muon, $|\psi _{+}\rangle $ and $|\psi _{-}\rangle $.
Hence, the total effective Hamiltonian is 
\begin{equation}
H_{eff}=\left( 
\begin{array}{cc}
E & i(k_{1}\frac{\omega _{2}}{2}-\mu B) \\ 
-i(k_{2}\frac{\omega _{2}}{2}-\mu B) & E
\end{array}
\right) ,
\end{equation}
where $\mu $ represents the total magnetic moment of the muon, $B$ is the
magnetic field, $k_{1},k_{2}$ reflects the different coupling of rotation to
the two helicity states.

Let us study in detail some properties of $H_{eff}$ . A biorthonormal
eigenbasis $\{|\psi _{1,2}\rangle ,|\phi _{1,2}\rangle \}$ of $H_{eff}$ is
given by

\begin{eqnarray*}
|\psi _{1,2}\rangle &=&\frac{1}{\sqrt{2}}[\pm i\chi ^{\frac{1}{2}}|\psi
_{+}\rangle +|\psi _{-}\rangle ], \\
|\phi _{1,2}\rangle &=&\frac{1}{\sqrt{2}}[\pm i\chi ^{-\frac{1}{2}}|\psi
_{+}\rangle +|\psi _{-}\rangle ],
\end{eqnarray*}
where $\chi =\frac{k_{1}\omega _{2}-2\mu B}{k_{2}\omega _{2}-2\mu B}.$ Its
eigenvalues are

\[
E_{1,2}=E\pm R, 
\]
where

\[
R=\sqrt{(k_{1}\frac{\omega _{2}}{2}-\mu B)(k_{2}\frac{\omega _{2}}{2}-\mu B)}%
, 
\]
therefore $E_{1,2}$ either are real or complex-conjugates. This peculiarity
of its spectrum ensures us that $H_{eff}$ is a pseudohermitian Hamiltonian 
\cite{mI}, and indeed an Hermitian operator $\eta $ exists which transform $%
H_{eff}$ into $H_{eff}^{\dagger }$ (see Eq.(1)). In the case of real
spectrum, for instance, $\eta $ assumes the form \cite{mI}\cite{so}:

\[
\eta =|\phi _{1}\rangle \langle \phi _{1}|+|\phi _{2}\rangle \langle \phi
_{2}|=\left( 
\begin{array}{cc}
\frac{1}{\chi } & 0 \\ 
0 & 1
\end{array}
\right) . 
\]

According to \cite{pa}, a violation of ($P$ and) $T$ in $H_{eff}$ would
arise though $k_{2}-k_{1}\neq 0$. We can improve the discussion on the $T$%
-violating parameters values, by means of the Theorem in Sect. 2 . Indeed $%
H_{eff}$ cannot be $T$-invariant for all the values of $k_{1}$ and $k_{2}$
which satisfy the condition 
\begin{equation}
(k_{1}\frac{\omega _{2}}{2}-\mu B)(k_{2}\frac{\omega _{2}}{2}-\mu B)>0
\end{equation}
since in this case $H_{eff}$ has a real, non degenerate spectrum. (Note that
by a suitable choice of $B$, condition (9) can be verified for all $%
k_{1},k_{2}.$)

Let us indeed evaluate the (non unitary) evolution operator $U(t)$. This is
given by \cite{fa}

\begin{eqnarray}
U(t) &=&|\psi _{1}\rangle e^{-iE_{1}t}\langle \phi _{1}|+|\psi _{2}\rangle
e^{-iE_{2}t}\langle \phi _{2}|= \\
&&\left( 
\begin{array}{cc}
e^{-iE_{1}t}+e^{-iE_{2}t} & i\chi ^{\frac{1}{2}}(e^{-iE_{1}t}-e^{-iE_{2}t})
\\ 
-i\chi ^{-\frac{1}{2}}(e^{-iE_{1}t}-e^{-iE_{2}t}) & e^{-iE_{1}t}+e^{-iE_{2}t}
\end{array}
\right) .  \nonumber
\end{eqnarray}

Then, assuming the initial condition $|\psi (0)\rangle =|\psi _{-}\rangle $,
the muon's state at the time $t$ is

\[
|\psi (t)\rangle =\frac{1}{2}[i\chi ^{\frac{1}{2}%
}(e^{-iE_{1}t}-e^{-iE_{2}t})|\psi _{+}\rangle
+(e^{-iE_{1}t}+e^{-iE_{2}t})|\psi _{-}\rangle ]. 
\]

The spin-flip probability is therefore

\begin{equation}
P(t)_{\psi _{-}\rightarrow \psi _{+}}=|\langle \psi _{+}|\psi (t)\rangle
|^{2}=\frac{\chi }{2}[1-\cos 2Rt],
\end{equation}
which agrees with the analogous calculation in \cite{pa} (where, however,
also the width $\Gamma $ of the muon is taken into account).

Note that the above probability do not depend on the sign of the time; this
feature occurs whenever (in a two level system) a transition probability
between orthogonal states is considered, and disappears when a different
choice of the states is made. Actually, evaluating for instance the
transition probability between the states $|\psi _{-}\rangle $ and $|\varphi
\rangle =\frac{1}{\sqrt{2}}[|\psi _{+}\rangle -|\psi _{-}\rangle ]$ one
obtains 
\begin{equation}
P(t)_{\psi _{-}\rightarrow \varphi }=|\langle \varphi |\psi (t)\rangle |^{2}=%
\frac{1}{2}(\cos Rt+\chi ^{\frac{1}{2}}\sin Rt)^{2},
\end{equation}
and $P(t)_{\psi _{-}\rightarrow \varphi }-P(-t)_{\psi _{-}\rightarrow
\varphi }=\chi ^{\frac{1}{2}}\sin 2Rt\neq 0$ , which explicitly shows that $%
H_{eff}$ is a $T$-violating Hamiltonian (even if $k_{1}=k_{2}$), in
agreement with our Theorem.

.

\end{document}